\documentclass[letterpaper,twocolumn,showpacs,preprintnumbers,amsmath,amssymb]{revtex4}

\usepackage{graphicx}
\usepackage{dcolumn}
\usepackage{bm}

\DeclareGraphicsExtensions{.eps,.EPS}

\begin{document}

\preprint{}

\title{Enhancement of Rydberg atom interactions using ac Stark shifts}


\author{P.~Bohlouli-Zanjani}
\author{J.~A.~Petrus}
\author{J.~D.~D.~Martin}
\affiliation{%
Department of Physics and Astronomy and 
Institute for Quantum Computing \\
University of Waterloo, Waterloo, Ontario, N2L 3G1, Canada
}%

\date{\today}

\begin{abstract}
The ac Stark effect was used to induce resonant energy transfer
between translationally
cold Rydberg atoms.  The $^{85}$Rb Rydberg atoms were obtained
by laser excitation of cold atoms from a magneto-optical trap.  
Using two-photon microwave spectroscopy of the $49s_{1/2}-50s_{1/2}$
transition, the field strength of a 28.5 GHz dressing field was
calibrated.  When this dressing field was set at specific
field strengths, the two-atom dipole-dipole process 
$43d_{5/2}$ + $43d_{5/2} \rightarrow 45p_{3/2}+41f$
was dramatically enhanced, due to induced degeneracy of the initial
and final states.  A series of observed resonant field strengths
correspond to
different magnetic sublevel possibilities for the initial and final
states.   These agree well with calculated resonance fields based
on a perturbative ac Stark shift formula. 
This method for enhancing interactions
is complementary to dc electric-field-induced resonant
energy transfer, but has more flexibility due to the possibility
of varying the applied frequency.  
\end{abstract}


\maketitle

The large transition dipole moments of Rydberg atoms make them much
more sensitive to electric fields than less excited atoms.  
For example, the dc
polarizabilities of low-angular momentum Rydberg states scale like
$n^7$, where $n$ is the principal quantum number \cite{gallagher:1994}.  
This high sensitivity can be exploited for various means.  For instance,
energy transfer between Rydberg atoms may often be
tuned into resonance using dc electric fields \cite{safinya:1981}.

Rydberg atoms are also sensitive to small oscillating electric
fields.   For example, microwave dressing
fields may be used to modify the dc polarizabilities of Rydberg
states \cite{hyafil:2004}.  
This could be used to reduce the influence of electric field 
inhomogeneities on the dephasing of a Rydberg atom qubit.  

The present work demonstrates the use of weak microwave
dressing fields to tune electric dipole-dipole
interactions between Rydberg atoms into resonance, 
in a case where this cannot
be accomplished by dc fields.  We exploit the latitude to change
both the amplitude {\em and} frequency of the dressing fields to
create interatomic interactions which are much stronger than could
otherwise be achieved.

Our results may be understood in terms of the ac Stark shifts of
the relevant states.
The ac Stark shift of a state $|\phi\!\!>$, in an 
electric field oscillating at angular frequency $\omega$,
with amplitude $\varepsilon_z$ in the $z$ direction,
is given by (see, for example, Ref.~\cite{haas:2006}):
\begin{equation}
\Delta E_{\phi} = \frac{1}{2} \varepsilon_z^2 
\sum_{m \ne \phi} \frac{(E_{\phi}-E_m) 
|\!\!<\!\! \phi | \mu_z | m \!\!>\!\!|^2}
{(E_{\phi}-E_m)^2 - (\hbar \omega)^2}
\label{eq:shift}
\end{equation}
where $E_m$ refers to the energy of state $|m \!\! >$
and $\mu_z$ is the electric dipole moment in the $z$ direction.
The modification of resonant energy transfer between Rydberg atoms by 
microwave fields has been previously studied 
in a strong field regime, where the Floquet description 
was more suitable \cite{kachru:1982,pillet:1983}.

\begin{figure}
\includegraphics{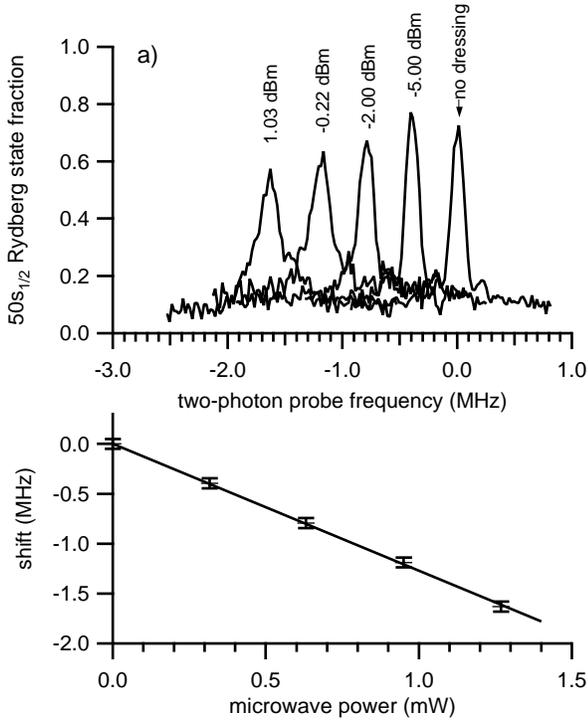}
\caption{\label{fg:shiftspectra}
(a) Observation of the two-photon $49s_{1/2}-50s_{1/2}$ microwave
transition for different powers of a 28.5 GHz dressing field.
The probe frequency shown on the horizontal axis is
twice the applied frequency, offset by $66.012926 \: {\rm GHz}$.
The synthesizer power settings are indicated.  (b) Shift
in the line centers $49s_{1/2}-50s_{1/2}$
as a function of synthesizer power.
Based on a calculated differential shift of 
$-0.195 \: {\rm GHz/(V/cm)^2}$ 
from Eq.~\ref{eq:shift},
we can find the calibration factor relating the synthesizer power
to $\varepsilon_z^2$.  
}
\end{figure}

The ac Stark shifts of Rydberg states can be probed using microwave 
spectroscopy.  For our purposes, 
this provides a useful check on the validity of Eq.~\ref{eq:shift},
and allows us to calibrate the applied field strengths when
studying interatomic interactions.
Our apparatus has been described previously 
\cite{afrousheh:2006,bohlouli:2006}.  
A standard vapor cell magneto-optical trap
(MOT) acts as a source of cold $^{85}$Rb atoms.  
These 
are excited to $49s_{1/2}$
Rydberg states using a $1 \: {\rm \mu s}$ pulse of laser light.  
Before excitation, the
magnetic and electric fields are reduced to less than $0.02 \: {\rm G}$ and
$0.05 \: {\rm V/cm}$ respectively.  Approximately $3 \: {\rm \mu s}$
after excitation, a $28.5 \: {\rm GHz}$ ac dressing
field is turned on.   While the dressing field is on, 
a $\approx 33 \: {\rm GHz}$ probe  
drives the $49s_{1/2}-50s_{1/2}$ two-photon transition.  
The probe pulse lasts $6 \: {\rm \mu s}$.  The
dressing field is then switched off and a selective field ionization
pulse \cite{gallagher:1994} is applied to measure the
$49s_{1/2}$ and $50s_{1/2}$ populations.
By scanning
the probe frequency between laser shots and collecting the
resulting spectra, we can measure the difference
in the ac Stark shifts of the two states involved
in the transition (see Fig.~\ref{fg:shiftspectra}).
As Eq.~\ref{eq:shift} indicates, the shifts should scale linearly
with applied dressing field power.  
This is verified in Fig.~\ref{fg:shiftspectra}b.

As shown in Fig.~\ref{fg:shiftspectra}, the linewidth
of the transitions increase as they shift.  By varying the
Rydberg density and probe power, we have determined that this is not
due to interatomic interactions or power broadening.  The observed
broadening is very sensitive to the alignment of the horn that
launches the dressing microwaves towards the atoms.  By varying the position
of this horn, the broadening can be made significantly worse
than shown in Fig.~\ref{fg:shiftspectra}.  Thus, we believe that
the broadening with increasing power is due to spatial inhomogeneity
of the dressing fields over the cold atom sample (this is complicated,
due to the presence of reflections and low-Q resonances within the
vacuum chamber).    Minimizing
this inhomogeneity is a major technical challenge in using
dressing fields.

By evaluating the matrix elements in Eq.~\ref{eq:shift} using
the techniques of Zimmermann {\it et al.}~\cite{zimmerman:1979},
we can compute the
frequency dependence of the ac Stark shift.  In particular,
as the denominator of Eq.~\ref{eq:shift} suggests, the shift direction
may be reversed by changing $\omega$.  
Spectroscopy probes the differential
shift between the two levels involved, and this may also be
reversed.
For example, we have observed that the 
$49s_{1/2}-50s_{1/2}$ transition is shifted to higher frequency
with a dressing field of 37.45 GHz, and that this shift is
also proportional to dressing power.
This ability to change the direction of differential
energy shifts with the applied frequency is essential to what 
follows.

\begin{figure}
\includegraphics{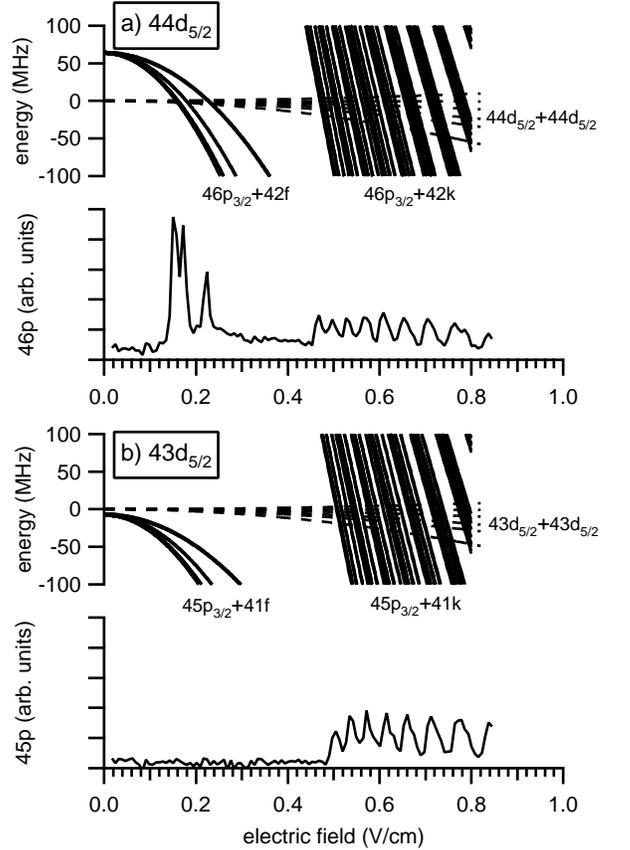}
\caption{\label{fg:ret}
DC electric-field-induced resonant energy transfer spectra, with 
atoms initially
in the a) $44d_{5/2}$ and b) $43d_{5/2}$ states.
Also shown are the calculated total energy of several two atoms 
states, relative to the zero field energy of 
$nd_{5/2}+nd_{5/2}$, where $n=44$ and $43$.  For the initial
states $nd_{5/2}+nd_{5/2}$ (dashed lines), we plot all magnetic 
sublevel possibilities for each of the two atoms 
involved ($m_{j}=1/2,3/2,5/2$).  In the case of the final states,
we consider the different possibilities for the two atom energies
depending on the magnetic sublevel of the $(n+2)p_{3/2}$ states
($m_j=1/2,3/2$), and the $(n-2)f$ and $(n-2)k$ states 
($m_j=1/2, ..., 7/2$).  The calculations follow the procedures of 
Ref.~\cite{zimmerman:1979}
and use the spectroscopic data of Ref.'s \cite{li:2003}
and \cite{han:2006}.
}
\end{figure}

It is well-known that dc Stark shifts may be used to enhance 
the interactions between Rydberg atoms \cite{safinya:1981,gallagher:1994}.
For example, consider the following resonant energy transfer process in Rb,
which may be driven by the dipole-dipole interaction:
\begin{equation}
nd_{5/2} + nd_{5/2} \rightarrow (n+2)p_{3/2} + (n-2)f_{5/2,7/2}.
\label{eq:process}
\end{equation}
For $n=44$ the energy of the final state is 
approximately 60 MHz higher
than that of the initial state (calculated using the
spectrosopic data of Ref.'s \cite{li:2003,han:2006}).
However, as
an electric field is applied, the $42f$ states exhibit strong,
quadratic Stark shifts to lower energies.
This tunes the process into resonance, as shown 
in Fig.~\ref{fg:ret}a.
In particular, atoms are excited to
$nd_{5/2}$ Rydberg states and allowed
to interact in the presence of a weak dc field.  By varying the
dc electric field between laser shots, and detecting the $(n+2)p$ 
population by selective field ionization, resonant energy transfer spectra
may be obtained 
(see Ref.~\cite{afrousheh2:2006} for more details).
As Fig.~\ref{fg:ret} illustrates, at $n=44$ the
resonance condition is turned on by a weak electric field, but
for $n=43$ it is tuned further out of resonance. In this
case the initial state is higher in energy than the final
state by $\approx 10 \: {\rm MHz}$.

\begin{figure}
\includegraphics{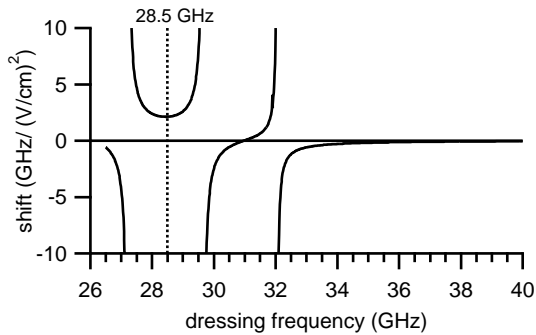}
\caption{\label{fg:ifcalc} Calculated difference in the ac Stark shifts
of the final and initial states of the process 
$43d_{5/2} + 43d_{5/2} \rightarrow 45p_{3/2} + 41f_{5/2}$
(all in $m_j=1/2$)
as a function of microwave frequency.  With no 
microwave field, the final state is lower in energy than the
initial state
by $6.0 \: {\rm MHz}$ and $8.3 \: {\rm MHz}$ 
(for $41f_{5/2}$ and $41f_{7/2}$ respectively)
\cite{li:2003,han:2006}.  Therefore, frequencies where the shift
is positive in this figure ({\it ie.} 28.5 GHz)
will push this process closer to resonance.
}
\end{figure}

Since dc fields cannot tune
$43d_{5/2} + 43d_{5/2} \rightarrow 45p_{3/2} + 41f_{5/2,7/2}$
into resonance, we consider using ac fields.
In particular, we expect that the flexibility in the choice
of $\omega$, which allows shift directions to be reversed, could
be beneficial.
To illustrate this, 
the difference in ac Stark shifts between the final
and initial states have been computed
as a function of frequency using Eq~\ref{eq:shift}.
Figure \ref{fg:ifcalc} illustrates that over certain frequency ranges
the ac field would shift the initial and final states closer into
resonance
-- which could not be achieved with a dc
electric field (see Fig.~\ref{fg:ret}b).

\begin{figure}
\includegraphics{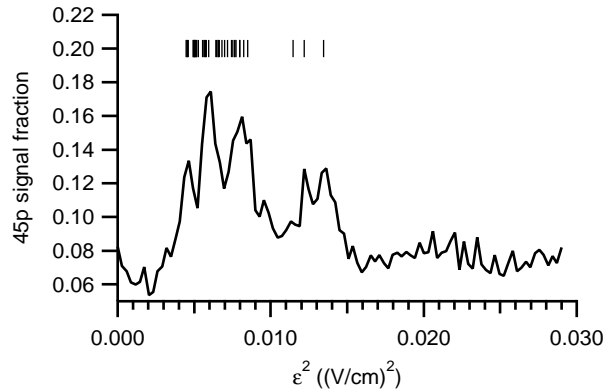}
\caption{
\label{fg:acret} 
Observed $45p$ signal as a fraction of total Rydberg
signal following a $23 \: {\rm \mu s}$ waiting period after
excitation of $43d_{5/2}$ Rydberg atoms,
with a 28.5 GHz field of variable strength present.  
The field amplitude calibration is discussed in the text.
The vertical lines indicate calculated resonance field strengths
for the different magnetic sublevel possibilities for
the initial and final states of 
$43d_{5/2} + 43d_{5/2} \rightarrow 45p_{3/2} + 41f_{5/2,7/2}$
(see text for details).
}
\end{figure}

To experimentally test this idea, we have looked for the resonant
energy transfer
process:
$43d_{5/2} + 43d_{5/2} \rightarrow 45p_{3/2} + 41f_{5/2,7/2}$,
in the presence of a 28.5 GHz microwave field.
The atoms are excited to 
$43d_{5/2}$ Rydberg states
(as discussed above), and the
dressing field is turned on $3 \: {\rm \mu s}$ after excitation. 
This field is held on for $20 \: {\rm \mu s}$ 
until shortly before selective
field ionization.   This is repeated, scanning the applied
microwave power between shots.  No deliberate dc 
electric field is applied.
As Fig.~\ref{fg:acret}
indicates, a significant population transfer to the $45p$ state is 
observed with the microwave field applied.  
As with the dc case, a series of resonances are observed.
This is expected, due to the different magnetic sublevel possibilities 
for the final and initial states (see below).
When the microwave
field strength is fixed at a resonance position and the overall
Rydberg density of the sample is varied (by changing the excitation
laser power), the fraction transferred decreases with reduced density.
This confirms that the observed resonances are due to an
interatomic process.
The transferred fraction grows linearly with density 
(for small transfer fractions, where depletion of the initial
states is not significant).
The widths of the resonances also increase with density.

To calibrate the applied fields in Fig.~\ref{fg:acret}, we observe
the single-photon transition $43d_{5/2}-45p_{3/2}$ with a series
of dressing field strengths.  
As with the $49s_{1/2}-50s_{1/2}$ case,
the shifts are observed to be linear in applied microwave power.
However, this case is slightly more complicated, due to the different
magnetic sublevels involved.  
By equating the experimentally observed shift 
of these transitions with those based on the prediction of 
Eq.~\ref{eq:shift}, we can determine 
the factor relating $\varepsilon_z^2$ to the applied microwave power.
This calibration can be compared with the field calibration
obtained from the $49s_{1/2}-50s_{1/2}$ line
discussed at the beginning
of this paper.  It is essential that these calibrations are done at the
same dressing field frequency, due to frequency
dependent losses, reflections and low-Q resonances 
within the vacuum chamber.  These calibrations agree to within 20 \%.  
Discrepencies may be due to the distribution of the 
probe field, which will have a slightly different spatial 
distribution over the trapped atom cloud at these two different 
frequencies.

Although the experiment is done with no deliberately applied
dc electric fields, they are difficult to avoid, since they
vary from day to day.  Therefore, the frequencies of each of
the $43d_{5/2}-45p_{3/2}$ and $43d_{5/2}-41f_{5/2,7/2}$
transitions are measured, with no dressing field present.
These give the experimental difference in energies
between the final and initial states of Eq.~\ref{eq:process}
for the field conditions of the experiment.
We find that this
``energy defect'', $\delta E \approx -7.4 \pm 0.1 \: {\rm MHz}$ for
the $41f_{5/2}$ case and $\approx -9.6 \pm 0.1 \: {\rm MHz}$ for
the $41f_{7/2}$ case.  These differ from the values of
$-6.0(5) \: {\rm MHz}$ and $-8.3(4) \: {\rm MHz}$
obtained from the constants of 
Ref.'s \cite{li:2003} and \cite{han:2006}.  
Although a small electric field explains some of this
discrepency, the shifts of the $43d_{5/2}-45p_{3/2}$ and 
$43d_{5/2}-41f_{5/2,7/2}$ lines are not consistent with 
the same field. 
In addition, the field required to explain the shift of the 
$43d_{5/2}-45p_{3/2}$ line exceeds the uncertainty in
the electric field zero ($\pm 0.05 \: {\rm V/cm}$), suggesting
that a slight adjustment of the spectroscopic constants may 
be necessary.

From the experimentally observed energy defects we can calculate
the resonance field strengths for the different magnetic sublevel
possibilities.
The ac Stark shifts for a state $|\phi\!\!>$ may be written
as $\Delta E_{\phi} = k_{\phi} \varepsilon_z^2$, where $k_{\phi}$ is 
computed using Eq.~\ref{eq:shift}.
For a process like Eq.~\ref{eq:process}
($|a\!\!>+|b\!\!> \: \rightarrow |c\!\!> + |d\!\!>$)
the resonance fields may be computed by rearrangement of:
$\delta E +  [k_c+k_d-k_a-k_b] \: \varepsilon_z^2 = 0$.
Due to the selection rules for the dipole-dipole interaction,
not all final and initial state magnetic sublevel combinations
are coupled.  In Fig.~\ref{fg:acret} the
vertical lines indicate all calculated resonance 
fields consistent with  $\Delta m_{j} =0,\pm 1$ for each atom.
Although many of the resonances are unresolved,
the general agreement is good, and the highest field resonances
are in clear agreement with the calculation.

Some caution is required in applying
Eq.~\ref{eq:shift} to the 41$f_{5/2,7/2}$ states, due to the
small energy separation of the two fine structure components
($2.3 \: {\rm MHz}$).  The fine structure
splittings of the other relevant states, $45p_{3/2,1/2}$ and
$43d_{3/2,5/2}$, are significantly larger.
When the Stark shifts become comparable
to the splitting of the two fine structure
components, Eq.~\ref{eq:shift}
is not valid.  This is entirely
analogous with the perturbative calculation of polarizabilities
in the dc case.   To examine this issue, a Floquet calculation
has been implemented (see, for example, Ref.~\cite{water:1990}). 
This calculation shows that
for the field strengths at the resonance
locations, the perturbative calculation is accurate on the scale
of Fig.~\ref{fg:acret}.  

The modification of resonant energy transfer between Rydberg atoms due
to {\em strong} microwave fields has been 
reported \cite{kachru:1982,pillet:1983}.  
In this case, the primary observation was
that an integer number of microwave photons can either
be given up or gained in resonant energy transfer to account for
the energy defect between the initial and final states.
Thus, the exact frequency of the ac field
plays an important role.  The microwave power determines the number
of ``sidebands'' present (the number of photons lost or gained
in the collision).  This can be accurately described using
Floquet theory \cite{pillet:1983} (although the ac Stark effect
does play a minor, observable role).
In the present work -- where the energy level 
shifts are perturbative -- the important tuning parameter is
the microwave power.  The frequency is not as important -- it
should be set within
a range to give the desired shift the correct sign.
However, 
it should not be too close to any resonance, as this will prevent
the microwave fields from being turned on and off adiabatically.

Rydberg atom interactions have recently received considerable attention
in the context of quantum information processing with neutral
atoms.  For example, the dipole-dipole interaction between Rydberg atoms 
has been proposed as means of 
allowing clouds of cold atoms to store qubits, using a process
known as dipole blockade \cite{lukin:2001}.
Lukin {\it et al.}~\cite{lukin:2001}
considered using long-range resonant electric dipole interactions.  
However, initial experiments in Rb have focused on non-resonant
van der Waals interactions \cite{tong:2004,singer:2004,liebisch:2005}.  
Recently, local blockade has been observed in Cs using 
resonant dipole-dipole interactions between 
Rydberg atoms \cite{vogt:2006}. 
In Rb, several groups have  identified Eq.~\ref{eq:process} 
as a strong resonant process 
\cite{walker:2005,reinhard:2006,reetz:2006}.
As the experimental results of the present work indicate, 
this process may be
shifted into resonance by either dc or ac electric fields
(Fig.'s \ref{fg:ret} and \ref{fg:acret}).  This would enhance
the blockade effect.  Possible advantages of ac fields over 
dc fields include the capacity to 
turn interactions on and off very quickly (due to the modulation
capabilities of the source), and the ability to induce
interactions at arbitrary dc fields.

In summary, perturbative ac fields have been demonstrated 
to enhance the interactions between Rydberg atoms by making them
{\em resonant}.  The frequency dependence of the ac Stark shift
allows this to be accomplished with more versatility than
the dc Stark shift.  

It is a pleasure to acknowledge useful discussions with J. D. Carter.
We thank K. Afrousheh for assistance in the early stages of this work.
This work was supported by NSERC, CFI, and OIT.

\end{document}